\documentclass[11pt]{article}
\usepackage{amsmath,amsbsy,amsfonts,amssymb,graphics,epsfig,float,mathrsfs,amsthm,braket}
\usepackage{ifsym}
\usepackage{psfrag}
\usepackage{color}
\usepackage[normalem]{ulem}

\newcommand{\upd}{\mathrm{d}}
\newcommand{\Bod}{\mbox{Bo}_d}
\newcommand{\Bo}{\mbox{Bo}}
\newcommand{\rhod}{\rho_d}
\newcommand{\rhol}{\rho_l}
\newcommand{\rhov}{\rho_v}%
\newcommand{\glv}{\gamma_{lv}}%
\newcommand{\gdl}{\gamma_{dl}}%
\newcommand{\gdv}{\gamma_{dv}}%
\newcommand{\Gdl}{\Gamma_{dl}}%
\newcommand{\Gdv}{\Gamma_{dv}}%
\newcommand{\lc}{\ell_c}%
\newcommand{\lcd}{\ell_c^{D}}%
\newcommand{\zmax}{z_{\mathrm{max}}}%
\newcommand{\rmax}{r_{\mathrm{max}}}

\newcommand{\s}[1]{{\textsf{\textbf{#1}}}}

\begin{document}
\title{\s{Non-wetting drops at liquid interfaces: From liquid marbles to Leidenfrost drops}}
\author{ \textsf{Clint Y. H. Wong$^\dagger$, Mokhtar Adda-Bedia$^\ast$ and Dominic Vella$^\dagger$}\\ 
{\it$^\dagger$Mathematical Institute, University of Oxford, UK}\\
{\it$^\ast$Univ. Lyon, ENS de Lyon, Univ. Claude Bernard, CNRS, Laboratoire de Physique,}\\{\it
F-69342 Lyon, France}
}

\date{\today}
\maketitle
\hrule\vskip 6pt
\begin{abstract}
We consider the flotation of deformable, non-wetting drops on a liquid interface. We consider the deflection of both the liquid interface and the droplet itself in response to the buoyancy forces, density difference and the various surface tensions within the system. Our results suggest new insight into a range of phenomena in which such drops occur, including Leidenfrost droplets and floating liquid marbles. In particular, we show that the floating state of liquid marbles is very sensitive to the tension of the particle-covered interface and suggest that this sensitivity may make such experiments a useful assay of the properties of these complex interfaces.
\end{abstract}
\vskip 6pt
\hrule

\maketitle

\section{Introduction}

Drops of one liquid at the surface of another are common at both large and small scales: oil slicks may break up into droplets \cite{Nissanka2017}, while  droplets of coffee on the surface of coffee are often observed, albeit briefly, when making a morning brew \cite{neitzel2002noncoalescence}.

Such droplets can only float in equilibrium when a number of different force balances are satisfied: their weight or buoyancy must be balanced by the net force from the underlying liquid. As with rigid particles \cite{vella2015floating}, the restoring force from surface tension can lead to the flotation of  droplets on the surface of a less dense liquid:  small drops of water may float on the surface of oil \cite{balchin1990capillary,phan2012can,phan2014stability}. Unlike rigid particles, however, such droplets may also deform greatly, forming either liquid lenses (when the droplet is partially wetting) or small (relatively non-wetting) droplets, close to spherical. Which of these possibilities occurs depends on the detailed balance between the three interfacial tensions at the contact line: the Neumann conditions \cite{Langmuir1933,de2013capillarity}.

The flotation of liquid lenses, particularly oil droplets  floating on a bath of water, has attracted the most attention.  To understand the shape of such lenses, previous work has focused on numerical approaches \cite{pujado1972sessile,boucher1980capillary,burton2010experimental}, although analytical results are available for very large, flat lenses \cite{pujado1972sessile}. 

For relatively non-wetting droplets, some analytical progress is possible when the droplet may be approximated as two spherical caps \cite{princen1965shape}. In other circumstances, particularly for larger drops, numerical techniques are again used \cite{elcrat2005numerical} though some understanding may be obtained by modelling the drop as two halves of an oblate spheroid \cite{ooi2015deformation}. Mahadevan \emph{et al.} \cite{Mahadevan2002} also presented a detailed study of the (low gravity) behaviour of a compound drop  sitting on a rigid substrate: here four phases frequently meet at a single point.

\begin{figure}[t!]
\centering
\includegraphics[width=0.5\columnwidth]{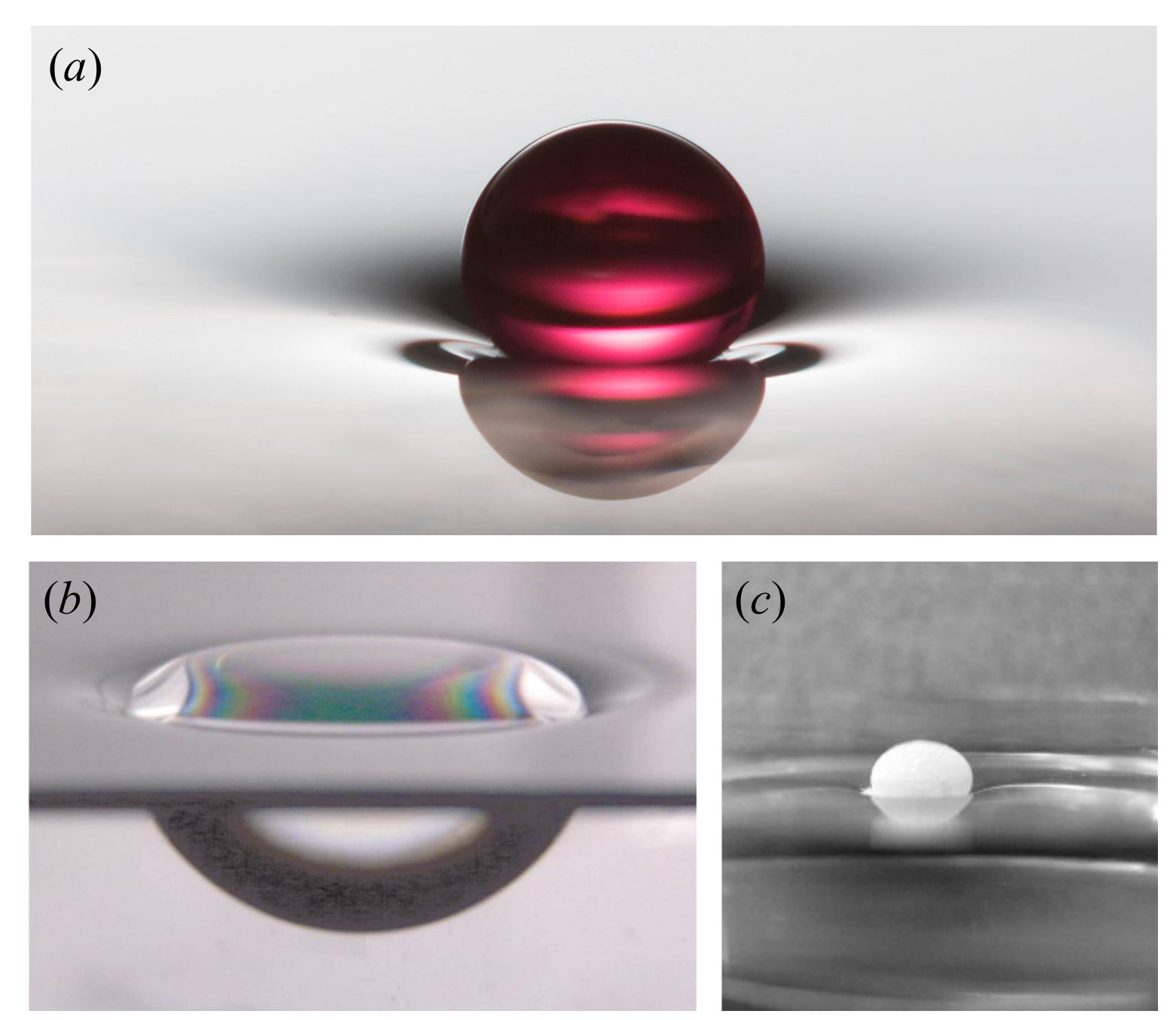}
\caption{Images of non-wetting drops `floating' on a liquid interface: (a) A dyed ethanol drop evaporating at the surface of a hot liquid pool (reproduced with permission from ref.~\cite{maquet2016leidenfrost}. Copyright (2016) by the American Physical Society), (b) a static drop consisting of the same fluid as the bath, separated  by a thin air film (reproduced with permission from ref.~\cite{couder2005bouncing}. Copyright (2005) by the American Physical Society) and (c) a liquid marble floating on water (reprinted from ref.~\cite{bormashenko2009water}, with permission from Elsevier). }
\label{Fig:Examples}
\end{figure}


Recent experiments have begun to focus on a range of problems that involve droplets floating at a liquid interface in what is close to, if not precisely, a non-wetting state.  For example, volatile drops levitate above a bath of warm liquid in the Leidenfrost state \cite{maquet2016leidenfrost}, see fig.~\ref{Fig:Examples}(\emph{a}), while the reverse scenario (a warm droplet levitating on a bath of liquid Nitrogen) has also been considered \cite{adda2016inverse}. A similar effect may be obtained by vibrating the liquid bath close to the Faraday threshold: the replenishing of the lubricating air film has the effect of maintaining the drops above a bath of the same liquid almost indefinitely \cite{couder2005bouncing,bush2015}, see fig.~\ref{Fig:Examples}(\emph{b}). Finally,  one of the most compelling demonstrations of the stabilizing properties of a particle coating is the ability of a particle coated droplet (a `liquid marble' \cite{aussillous2001liquid}) to float on the surface of the same liquid almost indefinitely, \cite{aussillous2001liquid,gao2007ionic,mchale2011liquid,Bormashenko2017} see fig.~\ref{Fig:Examples}(\emph{c}), even when motion occurs due to Marangoni flows \cite{Bormashenko2015}. Despite the importance of liquid marbles generally, previous analysis of floating liquid marbles has been primarily qualitative \cite{gao2007ionic,bormashenko2009mechanism}. Quantitative measurements on the dimensions of such drops have only been reported recently by Ooi \emph{et al.} \cite{ooi2015deformation,ooi2016floating}; this data has yet to be explained.

In this paper we seek to shed light on the manner in which a non-wetting (or very close to non-wetting) droplet floats at a liquid interface. In developing this insight we consider experimental data from a range of settings including Leidenfrost drops and liquid marbles and show that these can be understood as the appropriate limits of a single non-wetting drop model.

\section{Theoretical formulation}

\begin{figure}[h!] 
\centering
\includegraphics[width=.75\columnwidth]{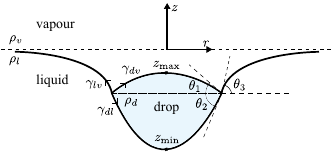}
\caption{Schematic diagram of a drop (blue) floating at a liquid--vapour interface. The three-phase contact line (TPCL) is the circle along which all three phases meet; the dotted horizontal line represents the projection of the TPCL onto the $(r,z)$ plane. }
\label{fgr:diagram}
\end{figure}

We consider an isolated axisymmetric drop, of density $\rho_d$ at the interface between an infinite liquid bath (of density $\rhol$) and vapour of negligible density, $\rhov\ll\rhol$ (see fig.~\ref{fgr:diagram} for a schematic of the setup). We assume that the three fluids are mutually immiscible and that there are three non-zero interfacial tensions as a result: the liquid--vapour, drop--liquid and drop--vapour interfacial tensions, which we denote by  $\glv$, $\gdl$ and $\gdv$, respectively.

\subsection{Governing equations}

\subsubsection{Laplace--Young equation}
When two immiscible fluids are in contact, the interfacial tension induces a discontinuity of pressure across the interface (see e.g.~ref.~\cite{de2013capillarity}):
\begin{equation}
[p]_-^+=\gamma\kappa
\label{L-Y}
\end{equation}
where the left-hand side denotes the pressure jump across the interface, $\gamma$ is the relevant interfacial tension and $\kappa$ is the total curvature of the interface.  Though simple, this equation allows us to determine the shapes of the three fluid-fluid interfaces since in equilibrium the pressure within each liquid is everywhere hydrostatic. 

For a deformable floating drop, there are three interfaces of interest, each of which satisfies a slightly different version of the Laplace--Young equation. We denote each of these three interfaces by an index $i$ (with $i=1$ denoting the drop--vapour interface, $i=2$ the drop--liquid interface and $i=3$ the liquid--vapour interface). We use an intrinsic parametrization of each interface  \cite{boucher1980capillary} using the arc-length $s$ and interfacial inclination $\phi_i(s)$; the interface shape may then be written $[r_i(s),z_i(s)]$. With this parametrization, simple trigonometry gives that 
\begin{eqnarray}
\frac{\upd r_i}{\upd s}&=\cos\phi_i, \label{rz1}\\
\frac{\upd z_i}{\upd s}&=\sin\phi_i, \label{rz2}
\end{eqnarray}
while the total curvature of each interface is \cite{boucher1980capillary}  
\begin{equation}
\kappa_i=\frac{\upd \phi_i}{\upd s}+\frac{\sin \phi_i}{r_i}\text{.}
\end{equation}
With this expression for $\kappa_i$, we may now equate the pressure change due to the interfacial tension, given by \eqref{L-Y}, to the hydrostatic pressure within each liquid. This leads to three different versions of the Laplace--Young equation, which each express the balance between hydrostatic pressure and pressure jumps due to surface tension. The simplest, and most familiar, of these is for the liquid--vapour meniscus; here atmospheric pressure is taken as the pressure datum and we find that
\begin{equation}
\rho_lgz_3=\gamma_{lv}\left(\frac{\upd\phi_3}{\upd s}+\frac{\sin\phi_3}{r_3}\right)\text{.}
\label{threedim}
\end{equation}
Within the drop, there is no universal pressure datum; we therefore introduce a constant pressure $p_0=\rho_dgz_0$, with $z_0$ some (\emph{a priori} unknown) vertical position. We find that the corresponding equations for $z_1$ and $z_2$ are 
\begin{equation}
\rho_d g(z_1-z_0)=\gamma_{dv}\left(\frac{\upd\phi_1}{\upd s}+\frac{\sin\phi_1}{r_1}\right)
\label{onedim}
\end{equation}
and
\begin{equation}
\rho_dg(z_0-z_2)+\rho_lgz_2=\gamma_{dl}\left(\frac{\upd\phi_2}{\upd s}+\frac{\sin\phi_2}{r_2}\right)\text{.}
\label{twodim}
\end{equation}

The equations \eqref{threedim}--\eqref{twodim}, together with \eqref{rz1} and \eqref{rz2}, must be solved with appropriate boundary conditions. For the outer meniscus, the conditions are that $z_3,\phi_3\to0$ as $s\to\infty$. Similarly, at the centre of the drop we have $\phi_1=\phi_2=r_1=r_2=0$.  The three-phase contact line (TPCL) requires greater thought. A number of further conditions emerge from continuity at the TPCL, since $r_1=r_2=r_3$ and $z_1=z_2=z_3$ there. It is also clear from fig.~\ref{fgr:diagram} that $\phi_1=-\theta_1$, $\phi_2=\theta_2$ and $\phi_3=\theta_3$ at the contact line. To determine the angles, $\theta_i$, however, we must consider in detail the equilibrium of the TPCL itself.

\subsubsection{Equilibrium at the contact line}

 For the drop to float in equilibrium, as illustrated in fig.~\ref{fgr:diagram}, the interfacial tensions, denoted by $\gamma_{lv}$, $\gamma_{dv}$ and $\gamma_{dl}$, have to balance at the TPCL. The three angles $\theta_1$, $\theta_2$ and $\theta_3$ must therefore satisfy the Neumann relations \cite{Langmuir1933,de2013capillarity}
\begin{eqnarray}
\theta_1+\theta_2&=&\pi-\cos^{-1}\left[\frac{\gamma_{dv}^2+\gamma_{dl}^2-\gamma_{lv}^2}
{2\gamma_{dv}\gamma_{dl}}\right],\label{neumann1}\\
\theta_1+\theta_3&=&\cos^{-1}\left[\frac{\gamma_{lv}^2+\gamma_{dv}^2-\gamma_{dl}^2}
{2\gamma_{lv}\gamma_{dv}}\right]\text{.}
\label{neumann2}
\end{eqnarray}

Notice that the Neumann conditions \eqref{neumann1}--\eqref{neumann2} only have solutions for certain combinations of the interfacial tensions: when the arguments of the inverse cosines is greater than unity in magnitude, the droplet must spread to form a layer, rather than forming a floating drop or lens \cite{Langmuir1933}. Here, we shall most often be interested in situations where the droplet is close to being perfectly non-wetting, so that $\theta_1+\theta_2\approx\pi$ and $\theta_2\approx\theta_3$. However, we maintain the general notation for the time being.

\subsection{Non-dimensionalization\label{sec:nondim}}

The balance between hydrostatic and capillary pressures, expressed in \eqref{threedim}, causes  interfacial deformations to decay over the \emph{capillary length}
\begin{equation}
\lc=\left(\frac{\glv}{\rhol g}\right)^{1/2}.
\end{equation} It is natural to use this length scale to non-dimensionalize lengths in this problem, introducing dimensionless variables $R_i=r_i/\lc$ etc. In performing this non-dimensionalization, three important dimensionless parameters emerge, namely
\begin{equation}
D=\frac{\rhod}{\rhol},\hspace{1cm}\Gdv=\frac{\gdv}{\glv},\hspace{1cm}\text{and}\hspace{1cm}\Gdl=\frac{\gdl}{\glv},
\label{eqn:NonDim}
\end{equation} which represent the ratio of densities of the droplet (relative to the bath liquid), as well as the ratios of the two other interfacial tensions, $\gdv$ and $\gdl$, measured relative to the liquid--vapour interfacial tension, respectively. Our primary interest is in relatively heavy droplets, $D\gtrsim1$.

The appearance of the various dimensionless parameters in \eqref{eqn:NonDim} may be attributed to the appearance of several different effective capillary lengths in the problem: whereas a drop on a rigid surface has a single, well-defined capillary length, in this problem both the capillary length of the bare interface, $\lc$, and the capillary length of the drop itself,
\begin{equation}
\lcd=\left(\frac{\gdv}{\rhod g}\right)^{1/2}=\left(\frac{\Gdv}{D}\right)^{1/2}\lc,
\end{equation} might also be of interest. We shall see that our numerical results may be interpreted physically using one or other of these two capillary lengths.

We also note that the two dimensionless interfacial tensions, $\Gdv$ and $\Gdl$, play a role in determining the contact angles, and hence the wettability of the droplet. For axisymmetric objects, this wettability plays a key role in determining an object's flotability \cite{vella2015floating}. The combination of angles $\theta_1+\theta_2$ and $\theta_1+\theta_3$ are completely determined by the values of $\Gdv$ and $\Gdl$. Our particular focus here shall be on perfectly non-wetting drops, for which we have the simple relation $\Gamma_{dl}=1+\Gamma_{dv}$, corresponding to $\theta_1+\theta_2=\theta_1+\theta_3=\pi$. However, there is still some freedom in the values of the angle themselves. For simplicity, we shall isolate this freedom in the angle $\theta_3$ (so that $\theta_1$ and $\theta_2$ are completely determined once $\theta_3$ is determined). 

The interfacial inclination of the outer meniscus at the TPCL, $\theta_3$, will depend on the precise conditions for the droplet floating at the interface. In particular, we expect  $\theta_3$ to be a function of the droplet volume $V$ (since larger, heavier drops will sink lower into the liquid, increasing $\theta_3$). We shall therefore treat $\theta_3$ as a function of the droplet volume, $V$. (Indeed Ooi \emph{et al.} \cite{ooi2016floating} plot experimental measurements of this relationship for floating liquid marbles.) The dimensionless droplet volume $V/\lc^3$ is therefore a key dimensionless parameter. However, it is more conventional to talk about the Bond number, $\Bo=r_0^2/\lc^2$, where $r_0=(3V/4\pi)^{1/3}$ is the radius of the drop if spherical. We therefore write
\begin{equation}
\Bo=\left(\frac{3V}{4\pi\lc^3}\right)^{2/3}
\end{equation} for the Bond number. We shall also find it helpful to discuss our results in terms of the Bond number using the droplet capillary length; we therefore introduce the droplet Bond number
\begin{equation}
\Bod=\frac{r_0^2}{(\lcd)^2}=\left[\frac{3V}{4\pi(\lcd)^3}\right]^{2/3}.
\label{eqn:Bod}
\end{equation}

\subsection{Numerical results}

\begin{figure}[t!]
\centering
\includegraphics[width=0.9\columnwidth]{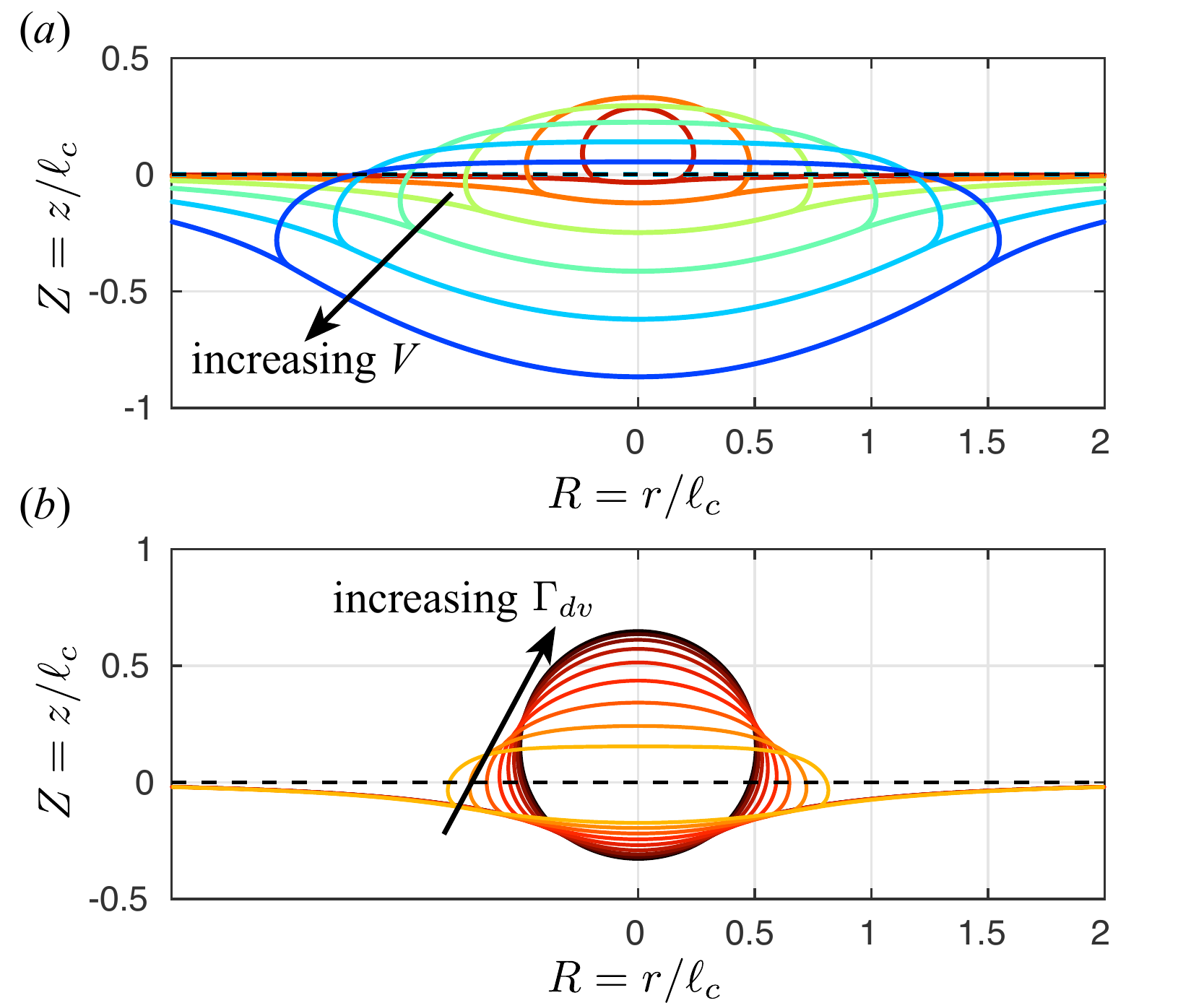}
\caption{Numerically determined droplet shapes for perfectly non-wetting drops,  so that $\Gdl=1+\Gdv$, with density ratio $D=2$.  (\textit{a}) The effect of increasing the droplet volume with $\Gdv=0.1$ held fixed throughout (so that $\Gdl=1.1$). Here, the angle $\theta_3$ is imposed and the corresponding volume computed. Results are shown with $\theta_3$ increasing from $5^{\circ}$ (red) to $30^{\circ}$ (dark blue), in increments of $5^{\circ}$, and corresponds to Bond numbers in the range $0.05\lesssim \Bo\lesssim 1.05$.  (\textit{b}) The effect of increasing the droplet--vapour interfacial tension, $\Gdv$, but maintaining a constant droplet volume, $\Bo=1/4$. Profiles are shown for $\Gdv=2^{i}$, $i=-5\ \text{(yellow)},-4,\dots,3 \ \text{(brown)}$, again with $\Gdl=1+\Gdv$.}
\label{fgr:transition}  
\end{figure}

We solve the dimensionless versions of equations \eqref{threedim}--\eqref{twodim} numerically (subject to appropriate boundary conditions) using MATLAB and Mathematica. The full dimensionless problem, together with details of the numerical scheme are given in Appendix A. From a computational point of view, it is simpler to impose a value of $\theta_3$ and calculate the droplet volume, or Bond number $\Bo$, that would give rise to this particular value of $\theta_3$. Figure \ref{fgr:transition}(\emph{a}) shows the droplet shapes obtained as $\theta_3$ varies. This increase of $\theta_3$ correspond to increasing Bond number. As expected, the drop flattens out as the Bond number grows --- gravity becomes more important.

The Bond number, $\Bo$, does not tell the whole story, however: fixing the value of $\Bo$ and altering the value of the droplet--vapour tension also changes the shape significantly, as shown in fig.~\ref{fgr:transition}(\emph{b}) for $\Bo=1/4$. Here we see that for moderate and large values of $\Gdv$ the droplets are essentially spherical (as should be expected since the Bond number is relatively small). However, decreasing $\Gdv$, the droplet becomes highly deformable and adopts the pancake configuration usually associated with large droplets \cite{de2013capillarity}.

Recalling that there are two capillary lengths in this problem, $\lc=(\glv/\rhol g)^{1/2}$ and $\lcd=(\gdv/\rhod g)^{1/2}$, there are therefore also two Bond numbers and it is natural to wonder whether the results shown in fig.~\ref{fgr:transition} can be understood solely  in terms of the droplet Bond number $\Bod$, defined in \eqref{eqn:Bod}.

\begin{figure}[t!]
\centering
\includegraphics[width=0.5\columnwidth]{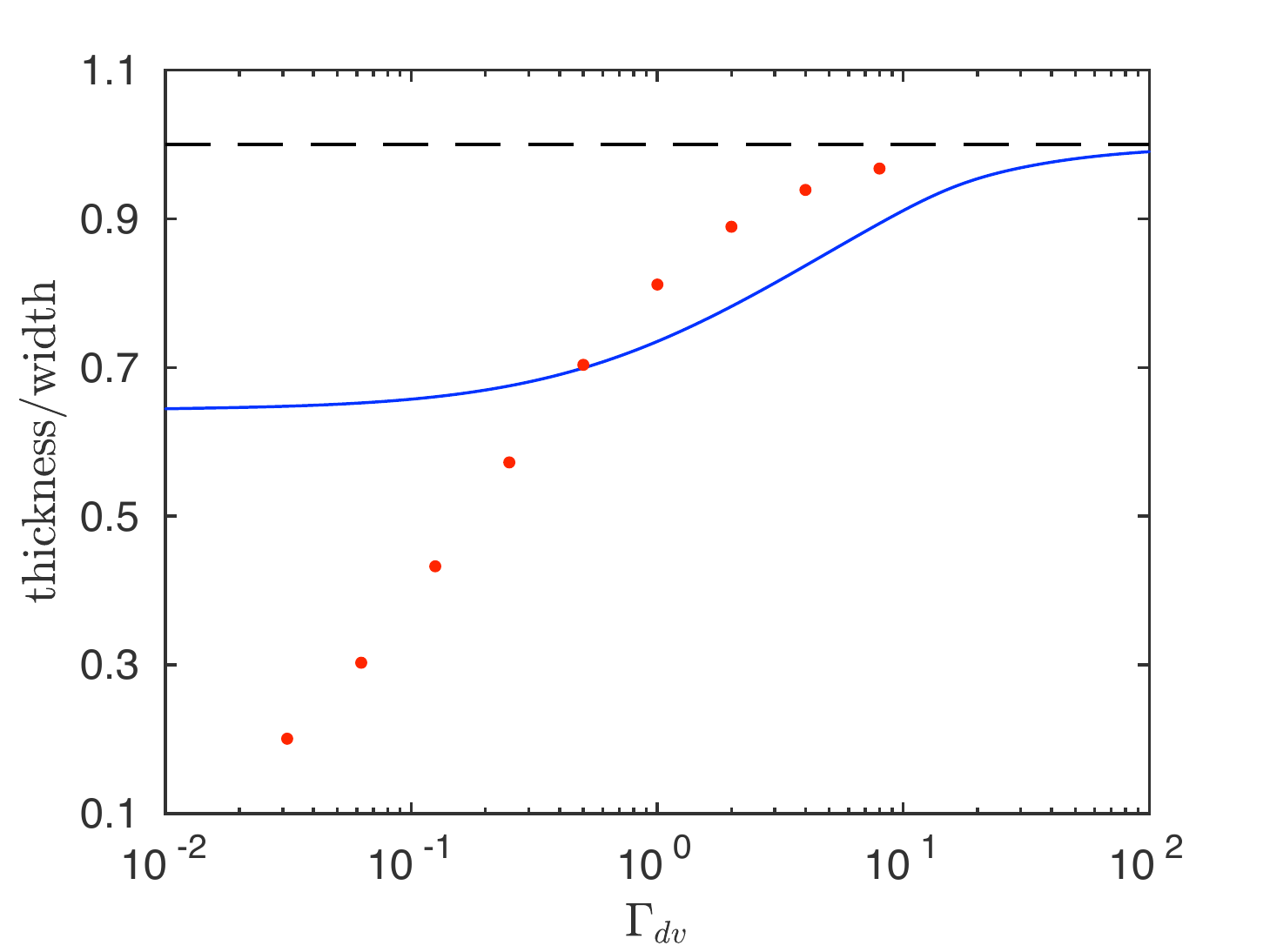}
  \caption{Plot of asymmetry (thickness/width) vs $\Gdv$ for perfectly non-wetting drops. The blue solid line indicates the situation where the droplet Bond number is fixed at $\Bod=[3V/(4\pi{(\lcd)}^3)]^{2/3}=1$, for $D=1$. The red dots correspond to results for the droplet profiles in fig.~\ref{fgr:transition}(\emph{b}). }
  \label{fgr:bond}
\end{figure}

The question is which of the two Bond numbers is the better description of how deformed a droplet is? Figure \ref{fgr:transition} shows that the drops become more deformed as they grow larger (as expected) and that with a fixed volume, but increasing $\Gdv$ the drop becomes more spherical. This suggests that the appropriate Bond number may be the droplet Bond number, i.e.~$\Bod$  is the relevant measure of a drop's deformability. Figure~\ref{fgr:bond} shows that this is also not the complete story: results with a fixed droplet Bond number,  $\Bod=1$, as well as droplet density, $D=1$, have an asymmetry that alters as the droplet--vapour tension changes. (Though, as expected the drop does revert to being approximately spherical  in the limit of very large $\Gdv$.)

From these numerical investigations it appears that, neglecting the effect of the droplet volume, the droplet--vapour tension $\Gdv$ gives a measure of the droplet's deformability. To understand this better, we now consider some analytical models of this problem, beginning with the case of `small drops', which we expect to be relatively undeformable.

\section{Relatively undeformable drops\label{sec:undeform}}

In the preceding section, we saw that there are a number of factors that affect the deformation of a floating drop: both its size (measured by the droplet Bond number $\Bod$) and the intrinsic value of $\Gdv$ can play a role. As such we now move on to consider the role of the droplet size and deformability separately. We begin by considering the case of small drops or, more accurately, droplets that deform the underlying liquid surface only slightly.

\subsection{Small non-wetting drops} 
We consider the limit of small substrate deformation, in the sense that $\theta_3 \ll 1$. In particular, for heavy drops $D\geq 1$, small deformations can only occur for drops with effective radius $r_0\ll \lc$. For the moment, we anticipate that small substrate deformations will correspond to small drop volumes but verify this \emph{a posteriori}.

In this situation, the droplet is approximately spherical, with radius $r_0\approx(3V/4\pi)^{1/3}$. This radius is determined by the volume of the droplet, and it, in turn,  determines the pressure within the drop, $p_d\approx 2\gdv/r_0$. However, the interface between the droplet and the liquid has a different interfacial tension, $\gdl=\gdv+\glv$, and, since the small size of the droplet ensures that the pressure within it is approximately constant, the drop--liquid interface must have radius of curvature
\begin{equation}
\tilde{r}_0=\frac{1+\Gdv}{\Gdv}r_0.
\end{equation} (We emphasize  here that to maintain a constant pressure within the drop, the drop must consist of two spherical caps of different radii joined together. Here our assumption is that the majority of the droplet volume is stored within the cap of radius $r_0$, with only a small perturbation from the cap that is contact with the substrate.)

The quantities of real interest, however, are the radial position of the contact line, $r_c$, and the interfacial inclination, $\theta_3$. Elementary geometry gives that $r_c=\tilde{r}_0\sin\theta_2=\tilde{r}_0\sin\theta_3$ (since, for non-wetting drops, $\theta_2=\theta_3$). We therefore have that
\begin{equation}
r_c=\frac{1+\Gdv}{\Gdv}r_0\sin\theta_3.
\label{rcro}
\end{equation}

The final ingredient required to close the system, and determine the behaviour of $r_c$ and $\theta_3$ as functions of the droplet volume $V$, is the global force balance. The restoring from the liquid is dominated by the force due to surface tension, since the object is small \cite{de2013capillarity,vella2015floating}, and so we find that the leading order vertical force balance is
\begin{equation}
\rho_d gV\approx 2\pi \gamma_{lv}r_c\sin \theta_3\approx2\pi \gamma_{lv}r_c\theta_3,
\label{stfb}
\end{equation} since $\theta_3\ll1$. Combining \eqref{stfb} with \eqref{rcro} we find that 
\begin{equation}
\theta_3\approx\sqrt{\frac{2}{3}}\left(\frac{D\Gdv}{1+\Gdv}\right)^{1/2}\Bo^{1/2}\ \propto\ \Bo^{1/2}
\label{theta3}
\end{equation}
while \eqref{rcro} immediately gives
\begin{equation}
R_c=\frac{r_c}{\lc}\approx \sqrt{\frac{2}{3}}\left[\frac{D(1+\Gdv)}{\Gdv}\right]^{1/2}\Bo \ \propto \Bo.
\label{Rc}
\end{equation} 
 The same theoretical prediction for the contact radius $r_c$ \eqref{Rc} in this regime has been reported by Princen \& Mason \cite{princen1965shape}. The scaling $r_c\sim \Bo$ has also been reported for non-wetting droplets on solid substrates \cite{mahadevan1999rolling,aussillous2001liquid,aussillous2006properties} and floating rigid spheres;\cite{vella2006load} however, the dependence on the interfacial tension $\Gdl$ and density $D$ are different here since they result from the deflection of three phases. (For example, for a rigid non-wetting sphere, $r_c/\lc\approx(2D/3)^{1/2}\Bo$.) The prediction for $\theta_3$ is, to our knowledge, novel.


\begin{figure}[t!]
\centering
\includegraphics[width=0.5\columnwidth]{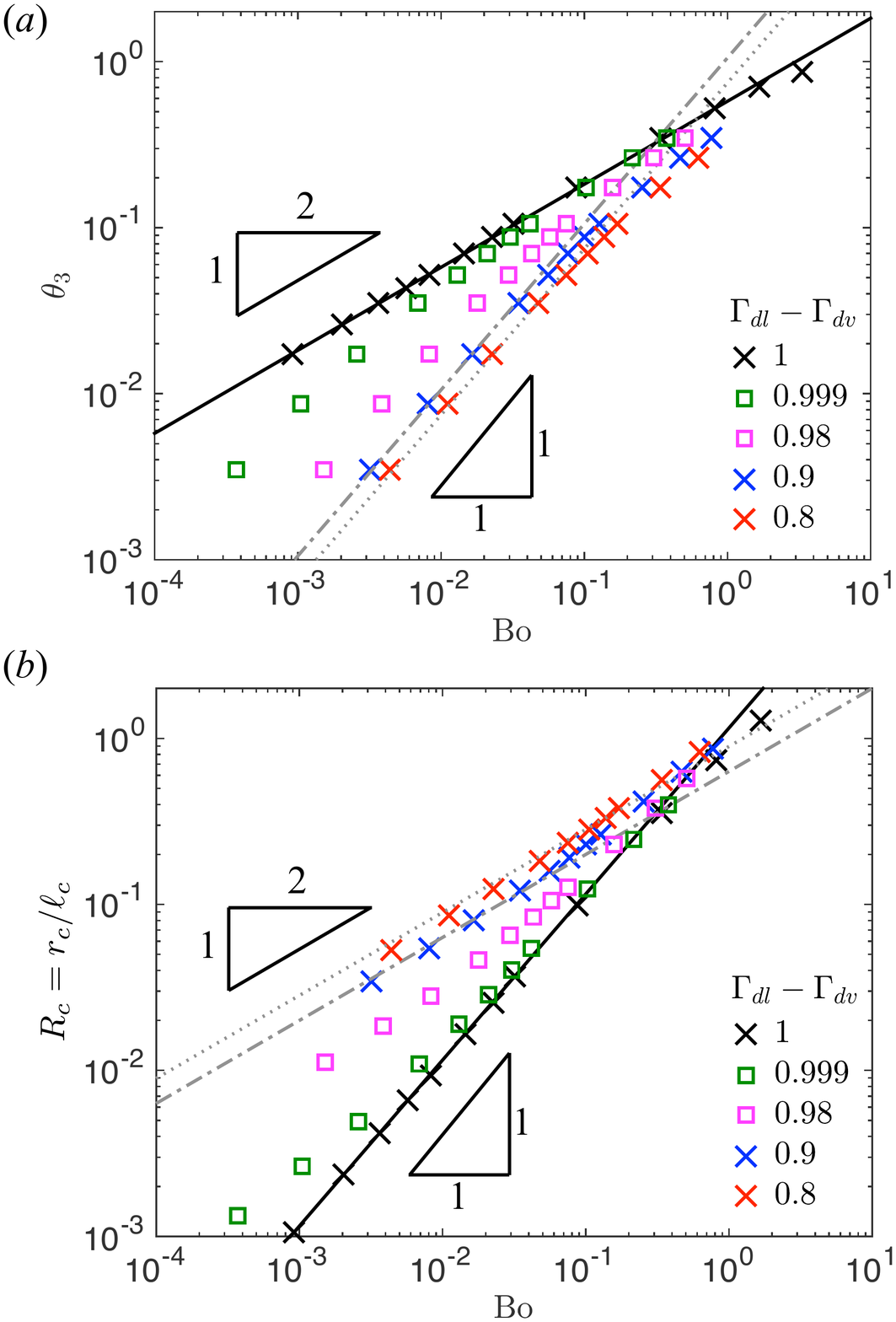}
  \caption{The transition of scaling behaviours for $\theta_3$ and $R_c$ respectively versus Bo as a droplet changes from non-wetting to partially-wetting. Shown is the case where $D=1$ and $\Gdv=1$. The black solid lines indicate the scaling predictions for the non-wetting case; the grey dash-dotted lines and dotted lines indicate the scaling predictions for $\Gdl=1.9$ and $\Gdl=1.8$ respectively. Intermediate cases are plotted with squares ($\square$).}
  \label{fgr:theta_Rc}
\end{figure}

\subsection{Small partially-wetting drops}

For perfectly non-wetting drops, the droplet angle, $\theta_1+\theta_2=\pi$. In reality, however, many super-hydrophobic materials are not `perfectly' non-wetting. For example, the droplet angle of liquid marbles has been reported  to be close to, but not exactly, $\pi$ \cite{ooi2016floating}. Here, we consider how the scaling laws  \eqref{theta3} and \eqref{Rc} are altered for hydrophobic droplets with $\Gdl-\Gdv<1$.

Numerical results for $\theta_3$ and $r_c/\lc$ as functions of $\Bo$ are shown in fig.~\ref{fgr:theta_Rc}.  Surprisingly, these numerical results show a significant deviation from the scaling laws \eqref{theta3} and \eqref{Rc}. Even with droplets that are very close to being perfectly hydrophobic, $\Gdl-\Gdv=0.999$, we observe significant deviations from \eqref{theta3} and \eqref{Rc} when the droplet Bond number becomes sufficiently small.

To understand this surprising behaviour, we reconsider the analysis of the last section. In particular, the vertical force balance, \eqref{stfb}, remains valid for the partially wetting case; the key difference, however, lies in the deduction of the contact radius, $r_c$ in terms of $r_0$ and the angle $\theta_3$. By equating capillary pressures, we have
\begin{equation}
r_c=\frac{\Gdl}{\Gdv}r_0\sin\theta_2
\label{rcro-partial}
\end{equation} which may be used to rewrite the vertical force balance \eqref{stfb} as 
\begin{equation}
2\pi\left(\frac{\Gdl}{\Gdv}r_0\sin\theta_2\right)\gamma_{lv}\sin\theta_3\approx\rho_d gV\text{.}
\label{stfb-partial}
\end{equation}

However, we cannot make the approximation $\sin\theta_2=\sin\theta_3\approx\theta_3$ here. Instead we must consider the Neumann conditions \eqref{neumann1}--\eqref{neumann2}. We are particularly interested in the effect of small changes in $\Gdl$ from the perfectly non-wetting value $1+\Gdv$. It is therefore natural to let
\begin{equation} 
\Gdl=(1+\Gdv)(1-\varepsilon)
\end{equation} and consider the behaviour for $\varepsilon\ll1$. From the Neumann conditions \eqref{neumann1}--\eqref{neumann2} we find that
\begin{align}
\theta_1+\theta_2&\approx\pi-\sqrt{2}\left(\frac{\varepsilon}{\Gdv}\right)^{1/2},\nonumber\\
 \theta_1+\theta_3&\approx\pi-\sqrt{2}\left(\frac{\varepsilon}{\Gdv}\right)^{1/2}(1+\Gdv)\nonumber
\end{align} and hence
\begin{equation}
\theta_2\approx\theta_3+\sqrt{2}\Gdv^{1/2}\varepsilon^{1/2}.
\label{eqn:Theta2Asy}
\end{equation}

Equation \eqref{eqn:Theta2Asy} gives some insight into the cause of the different behaviour that is observed when $\epsilon>0$ (i.e.~when the droplet is slightly wetting): as the interfacial deformation $\theta_3$ decreases (corresponding to smaller and smaller droplets), the angle $\theta_2$ saturates at a finite value that is set by the surface tensions in the problem. This is different to the perfectly non-wetting case, when $\theta_2$ decreases with $\theta_3$ without bound. For  $\sqrt{2}\Gdv^{1/2}\varepsilon^{1/2}\ll\theta_3\ll1$, corresponding to moderately small droplets, we shall recover the scalings of the perfectly non-wetting case, \eqref{theta3}--\eqref{Rc}. However, if instead $\theta_3\ll\sqrt{2}\Gdv^{1/2}\varepsilon^{1/2}\ll1$, corresponding to extremely small droplets, then we have the new scalings
\begin{equation}
\theta_3\approx\frac{\sqrt{2}}{3}D\left[\frac{\Gdv}{(1+\Gdv)(1+\Gdv-\Gdl)}\right]^{1/2}\Bo\propto\Bo
\label{theta3-partial}
\end{equation} and
\begin{equation}
R_c\approx\sqrt{2}\left[\frac{(1+\Gdv)(1+\Gdv-\Gdl)}{\Gdv}\right]^{1/2}\Bo^{1/2}\propto\Bo^{1/2}.
\label{Rc-partial}
\end{equation} Note that in both the non-wetting and partially wetting cases, the product $R_c\theta_3$ is fixed by vertical force balance to take the value $R_c\theta_3\approx\tfrac{2}{3}D~\Bo^{3/2}$, provided that $\theta_3\ll1$.

 The new scaling behaviours ($\theta_3\sim\Bo$, rather than $\theta_3\sim\Bo^{1/2}$ in the non-wetting case and vice versa for $R_c$) are verified by comparison with numerical results in fig.~\ref{fgr:theta_Rc}. We note that for a given value of $\varepsilon$ (i.e.~fixed values of $\Gdv$ and $\Gdl$) there is a smooth transition between the two sets of scaling laws as the droplet size, measured via the Bond number  $\Bo$, varies. Furthermore, as $\varepsilon\to0$, i.e.~as $\Gdl\to1+\Gdv$, the transition between scalings occurs for smaller and smaller droplets: as $\varepsilon$ decreases, \eqref{eqn:Theta2Asy} makes it clear that only the very smallest drops will be affected by the effects of partial wetting, while other drops will behave as non-wetting drops to all intents and purposes.

\subsection{Floating liquid marbles}

In recent years, a very striking demonstration of super-hydrophobicity has been the formation of so-called `liquid marbles': droplets of aqueous liquid encapsulated by a super-hydrophobic powder \cite{aussillous2001liquid,aussillous2006properties,mchale2011liquid,daisuke2015xray}. While liquid marbles are often encountered on solid, effectively rigid, surfaces, one of the most striking features is that simply by coating a droplet of water with hydrophobic grains, the droplet is able to sit on the surface of water without coalescing (see fig.~\ref{Fig:Examples}(\emph{c})). This coalescence may even be delayed for several weeks \cite{gao2007ionic}. However, the effective surface tension coefficient of such marbles has been somewhat controversial. A variety of techniques have been proposed to measure this tension, including analysis of the shape of a droplet sitting on a rigid surface \cite{aussillous2006properties}, as well as direct measurement of the capillary pressure within the droplet \cite{arbatan2011measurement}. 

\begin{figure}[t!]
\centering
\includegraphics[width=0.8\columnwidth]{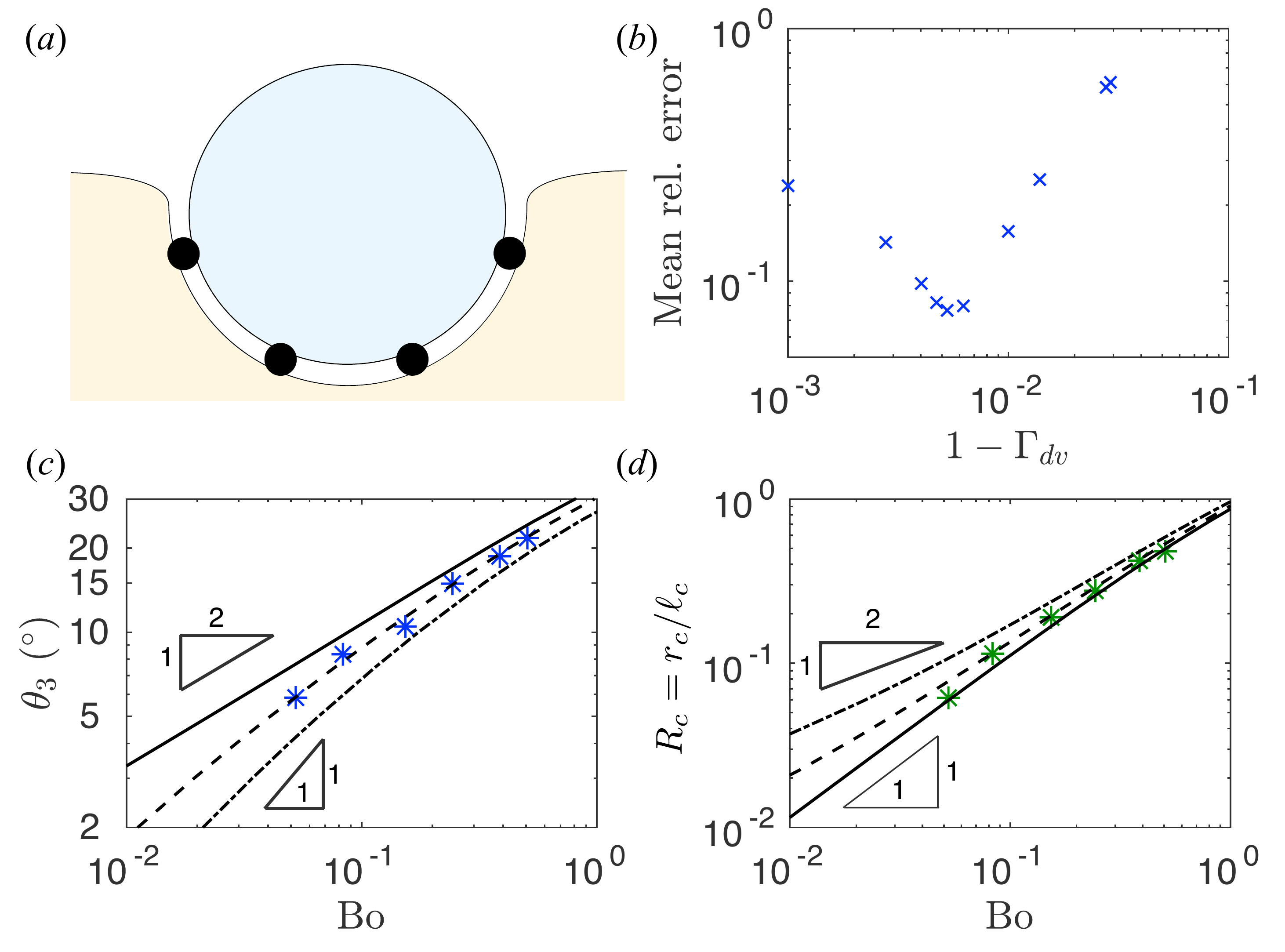}
\caption{Explaining the experimental data from ref.~\cite{ooi2016floating} with the liquid bridging postulate. (\textit{a}) Possible configuration of the drop-liquid interface of a liquid marble: the coating agglomerates (solid circles) are bridging between the drop and the substrate. Note that the thickness of the air gap and the size of the agglomerates have been exaggerated. (\textit{b}) Mean relative error in data fitting of $\theta_3$ for different values of $\Gdv$. (\textit{c})-(\textit{d}) Results of numerical simulations for $\theta_3$ and $r_c$ as functions of the Bond number $\Bo$ for different values of $\Gdv$: $\Gdv=0.971$ (dash-dotted), $\Gdv=0.995$ (dashed), and $\Gdv=1$ (solid). Asterisks are used to indicate the experimental data of ref.~\cite{ooi2016floating}. }
\label{fgr:Marbles}
\end{figure}

In this section, we have seen that even small departures from being perfectly non-wetting can have surprisingly large effects on the behaviour of  floating drops, as measured in the interfacial inclination and the radius of contact. Recent experiments \cite{ooi2016floating} on floating liquid marbles report measurements of both $\theta_3$ and $R_c$ as functions of the Bond number. These experimental results (as determined from digitization of fig.~5 of ref.~\cite{ooi2016floating}) are shown in fig.~\ref{fgr:Marbles}(\emph{c},\emph{d}) and show that the droplets do not obey simple power-law scalings. It is then natural to wonder whether these deviations encode useful information about the state in which liquid marbles float?

\subsubsection{The effective interfacial tension}

Ooi \emph{et al.} \cite{ooi2016floating} also present experimental data showing that the marble contact angle, $\theta_1+\theta_2$ is slightly less than $\pi$ (in particular, they found that $\theta_1+\theta_2\approx170^\circ$). This shows that the droplet is not in a perfectly non-wetting state, and hence that $\Gdl\neq1+\Gdv$ in our notation. Instead of levitating, as would be required for the liquid marble to be truly non-wetting, we therefore assume that the hydrophobic grains `bridge' both liquid gas interfaces (as shown in fig.~\ref{fgr:Marbles}(\emph{a})).  This is consistent with the fact that it is energetically favourable for the particles to absorb at any liquid interface \cite{mchale2011liquid,binks2002particles}. This assumption also retains the symmetry that the two liquids are identical, though we note that the two liquid--vapour interfaces  must remain out of contact for the liquid marble to sit stably at the interface. In this configuration then we expect that the effective tension of the drop--liquid interface is
\begin{equation}
\Gdl=\Gdv+\Gdv=2\Gdv.
\end{equation}

The value of $\Gdv$ (corresponding to the tension of a single particle coated interface) has been reported to take several different values depending on the liquids used, as well as the type and size of grains used.  \cite{Bormashenko2013} The first experiments, performed by Aussillous \& Qu\'{e}r\'{e} \cite{aussillous2006properties}, reported a value of $\Gdv\approx1$ for water droplets coated with silica, while glycerol and water droplets coated with lycopodium gave values of $\Gdv\approx0.70$ and $\Gdv\approx0.71$, respectively. The experiments of Ooi \emph{et al.} \cite{ooi2015deformation}, using micron-scale polytetrafluoroethylene (PTFE) powder on water, gave, for a single interface, $\Gdv=0.944$. An independent investigation \cite{arbatan2011measurement} reported similar results using the same material ($\Gdv=1$ and $\Gdv=0.971\pm 0.008$, using two distinct methods of measurement). (We report dimensionless values here to avoid confusion from variations in $\glv$.) However, we note that despite small disagreements, and with the exception of lycopodium used by Aussillous \& Qu\'{e}r\'{e} \cite{aussillous2006properties}, all previous measurements of $\Gdv$ suggest that the coating induces only a small reduction to the original (water-vapour) surface tension. We therefore consider the effect of a small correction from unity, i.e.~we set $\Gdv=1-\delta$ with $\delta\ll1$. Expanding the Neumann conditions for $\delta\ll1$ and using $\Gdl=2\Gdv$, one easily finds
\begin{equation}
\theta_1+\theta_2\approx \pi-(1-\Gdv)^{1/2}
\end{equation} and hence
\begin{equation}
\Gdv\approx1-(\pi-\theta_1-\theta_2)^2.
\label{eqn:GdvMarAng}
\end{equation} 

\subsubsection{Comparison with experimental results}

Based on \eqref{eqn:GdvMarAng}, it is tempting to try to infer the value of $\Gdv$ from measurements of the marble angle $\theta_1+\theta_2$. Using the range of values measured by Ooi \emph{et al.} \cite{ooi2016floating}, namely $165^\circ\lesssim\theta_1+\theta_2\lesssim175^\circ$, we find that $0.931\lesssim\Gdv\lesssim0.992$. While these values are consistent with previous measurements, they do not constrain the value of $\Gdv$ particularly closely.

An alternative approach is to use the experimental measurements of $\theta_3$ as a function of $\Bo$ (see fig.~\ref{fgr:Marbles}(\emph{c})), using $\Gdv$ as a single fitting parameter and constraining $\Gdl=2\Gdv$.  Figure \ref{fgr:Marbles}(\emph{b})  shows the mean relative error between experiments (measured for six different volumes by Ooi \emph{et al.} \cite{ooi2016floating}) and our numerical results for different values of $\Gdv$. In particular, since our numerical scheme computes the volume by fixing $\theta_3$, we define this relative error to be the relative discrepancy between the experimental volume and the computed volume, for a given $\theta_3$.  In fig.~\ref{fgr:Marbles}(\emph{b}), we observe that the mean relative error varies continuously with $\Gdv$ but has a minimum value at $\Gdv\approx 0.995$. 

Using the fitted value $\Gdv\approx0.995$ we may also compute the variation of $R_c$ with $\Bo$ (see fig.~\ref{fgr:Marbles}(\emph{d})). This gives an independent verification of our fitting (which was performed only using the interfacial inclination, $\theta_3$). We also note that $\theta_3(\Bo)$ is a very sensitive function of $\Gdv$: the theoretical results with $\Gdv=1$ and $\Gdv=0.971$, which corresponds to $\theta_1+\theta_2=180^{\circ}$ and $\theta_1+\theta_2=170^{\circ}$ respectively, provide a noticeable error when compared with experimental data. Finally, we note that for $\Gdv=0.995$, $\theta_1+\theta_2\approx 176^\circ$, which is just outside the range of values reported experimentally.

In summary, our investigation of the results of our model adapted to liquid marbles suggests that measurements of $\theta_3$ and $R_c$ as functions of marble size,  $\Bo$, provide a relatively sensitive way of estimating $\Gdv$. This sensitivity is associated with the difference in scalings of $\theta_3$ and $R_c$ with $\Bo$ depending on whether $\Gdv=1$ or not: the transition between the $\Bo^{1/2}$ and $\Bo$ scalings occurs gradually and encodes a good deal of information about the precise interfacial tensions involved. We also note that the value of $\Gdv$ we obtain through this procedure is consistent with previous estimates but suggests that experimental errors in those fitting procedures gives a relatively large uncertainty in the inferred value of $\Gdv$. 

We note that in the above calculation, we have assumed that the interface of liquid marbles behaves purely as that of a liquid drop. In fact,  particle--coated interfaces  have an elastic character \cite{Vella2004} consistent with a bending rigidity $B\sim\glv d^2$ where $d$ is the particle diameter. This elastic contribution can be neglected in comparison with the pressure due to surface tension, provided that $d/r_0\ll1$. Nevertheless, differences in particle packings may explain some of the variability in the effective surface tension coefficient $\Gdv$ reported previously.

\section{Deformable drops}

The analysis of the previous section relied on the assumption that the drops are relatively undeformable. This may be realized with a combination of large $\Gdv$ and/or small droplet Bond number. However, the alternative limit of relatively small $\Gdv$ and/or large droplet Bond number is also of interest. For example Leidenfrost droplets, which are rendered non-wetting by a layer of vapour, may `float' at a liquid interface  \cite{adda2016inverse,maquet2016leidenfrost} while a vibrating bath may support a long-lived droplet of the same fluid at the interface \cite{couder2005bouncing,bush2015} (see fig.~\ref{Fig:Examples}(\emph{b})). In such scenarios, the surface tension of the liquid droplet is usually the same (or similar to) that of the liquid bath, so that $\Gdv\approx1$. (One exception to this is water drops performing the Leidenfrost effect on a bath of liquid Nitrogen, for which $\Gdv\approx8$.\cite{adda2016inverse}) We begin by considering recent experimental work on Leidenfrost droplets \cite{maquet2016leidenfrost}, which requires numerical analysis. We then move on to consider the case of much larger droplets, which form `puddles' analogous to those observed on solid substrates \cite{aussillous2006properties}, and can be understood analytically.

\subsection{Leidenfrost drops}

Leidenfrost drops  levitate due to the evaporation of the droplet --- the resulting vapour produces a thin cushioning layer that prevents the droplet from touching the substrate \cite{quere2013leidenfrost}. In general the temperature of the substrate must exceed the droplet's boiling point by a great deal to allow for this levitation. However, drops may levitate on a heated liquid bath even when the temperature of the bath exceeds the droplet's boiling point only slightly since the surface in such cases is very smooth \cite{maquet2016leidenfrost}, as shown in fig.~\ref{Fig:Examples}(\emph{a}). Determining the shape of the levitating droplet in such circumstances is generally complicated since the thickness of the vapour layer varies spatially. Such a calculation has been performed for Leidenfrost droplets on a rigid substrate \cite{Snoeijer2009,Maquet2015} but only for floating Leidenfrost droplets in the limit of small interfacial deformations \cite{maquet2016leidenfrost}. Here we show that the model of non-wetting droplets developed above is able to explain features of experimental results  obtained with larger interfacial deformations.

\begin{figure}[h!]
\centering
\includegraphics[width=0.6\columnwidth]{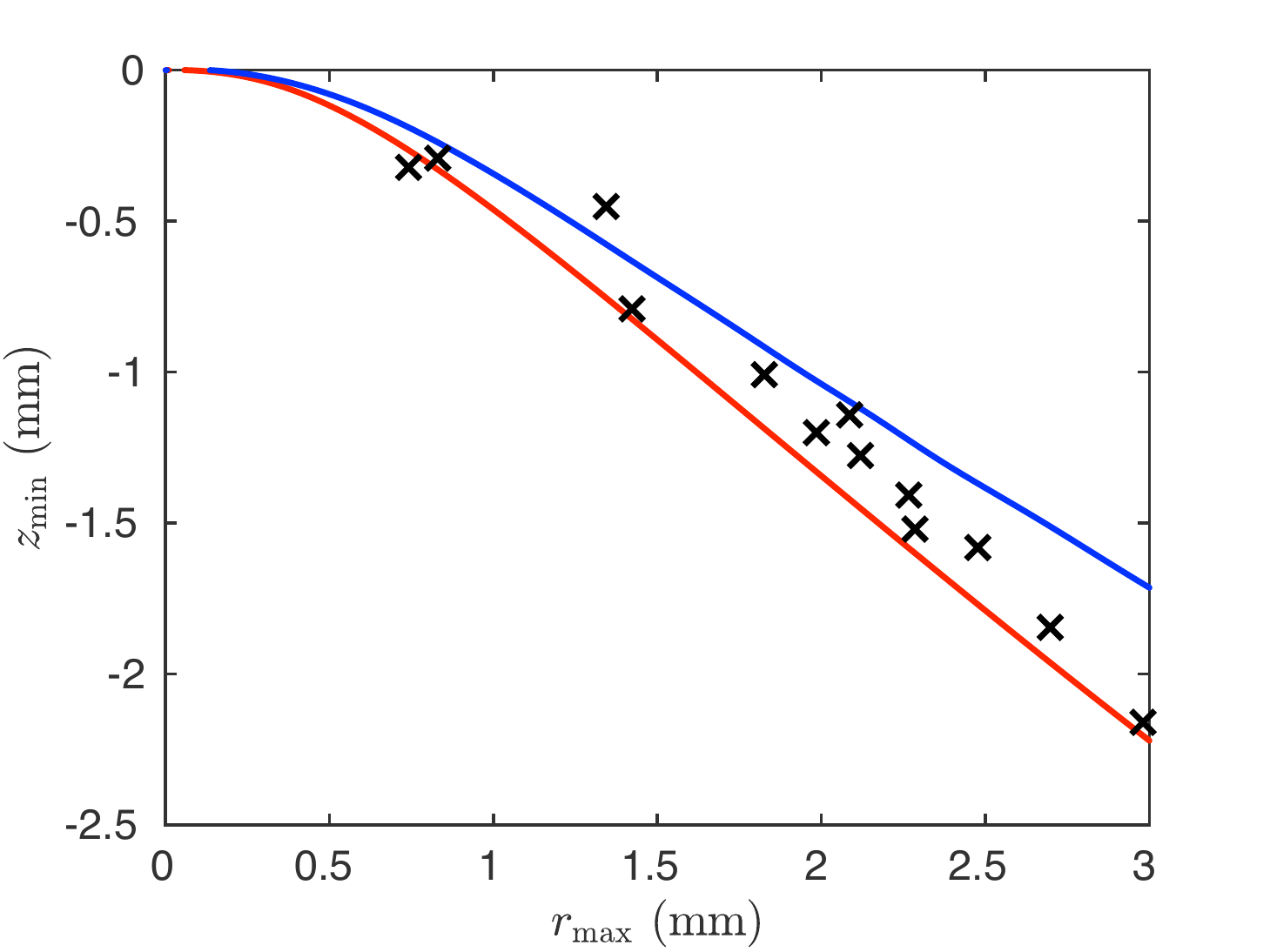}
\caption{Experimental results for the maximum vertical displacement of the liquid substrate for a Leidenfrost drop of maximum radius $r_\text{max}$ ($\times$) \cite{maquet2016leidenfrost}. The blue (dashed) curve gives the theoretical prediction by Maquet \emph{et al.} \cite{maquet2016leidenfrost} and the red (solid) curve gives the theoretical prediction with the non-wetting drop model of \S2.}
\label{fgr:maquet_theory}
\end{figure}

Maquet \emph{et al.} \cite{maquet2016leidenfrost} present data for ethanol drops levitating on a bath of heated silicone oil. In particular, they measure the depth of the base of the droplet below the undeformed interface ($z_{\mathrm{min}}$ in the notation of fig.~\ref{fgr:diagram}) as a function of the maximum radius of the droplet. These data are shown in fig.~\ref{fgr:maquet_theory}, together with the results of the detailed (but small-slope) lubrication model presented there. We note that for relatively large drops the discrepancy between the theory and experiment grows, which Maquet \emph{et al.} \cite{maquet2016leidenfrost} attribute to the growing interface deflections as the drops become larger.

Here we seek to understand this discrepancy by using our models of non-wetting drops: we do not take into account the details of the vapour layer, but merely assume that the droplet is non-wetting. (This corresponds to assuming that the vapour layer is very thin in comparison to the size of the droplet.) Ethanol drops on a silicone oil pool have $\Gdv\approx1.34$ (using measurements of the capillary length from fig.~3 of ref.~\cite{maquet2016leidenfrost}), and hence $\Gdl=1+\Gdv\approx2.34$. Using these values and computing the value of $z_{\mathrm{min}}$ numerically, we find good agreement between experimental data and the simple theory (see fig.~\ref{fgr:maquet_theory}). However,  we note that it does over-estimate the depth of the drops at a particular value of $\rmax$. This discrepancy may be the result of rounding errors in inferring the value of $\Gdv$ or, alternatively, a small shear from the Leidenfrost layer that acts to spread the drop further (and hence make it less deep for a given droplet volume). We conclude that Maquet \emph{et al.} \cite{maquet2016leidenfrost} are correct that large interface deformations become important for large droplets; we note, further, that the non-wetting droplet model may be used to provide a reasonable approximation to the shape of such drops.

\subsection{Very deformable drops}

The results above are purely numerical. However, it is possible to obtain some analytical understanding of the behaviour of very deformable non-wetting droplets (so that $\Gdl=1+\Gdv$). We are not aware of experimental data in this limit, but note that droplets that bounce on a vibrating bath may be subject to large deformations (since the accelerations, and hence effective gravity, are large) \cite{couder2005bouncing,Pucci2011}.

The analysis of relatively undeformable drops in \S\ref{sec:undeform} exploited the knowledge that in this limit the drop should remain approximately spherical. However,  the richness of this problem lies in the ability of all three phases to deform simultaneously. To obtain a more complete picture of our problem we therefore focus now in the opposite regime: that in which the drop is very deformable. 

It is well known that large deformable drops form `puddles'  or `pancakes' ---  drops flattened by gravity to have a thickness comparable to the capillary length \cite{aussillous2006properties}. In this work  we are primarily interested in capillary effects and it therefore seems that one cannot have droplets that are highly deformable, but supported primarily by surface tension (since $D>1$). This apparent contradiction is resolved by recalling that the floating droplet problem contains multiple capillary lengths. In particular, for heavy drops, it is possible for gravity to dominate the drop shape ($\Bod\gg1$), whilst simultaneously surface tension provides sufficient restoring force for the drop to remain afloat, i.e.~$\Bo\lesssim1$.

Recalling the numerical findings presented in fig.~\ref{fgr:transition}(\emph{a}) we expect that as $r_0$ increases, gravity becomes more important, encouraging the drop to spread and deform the substrate to minimize its (gravitational) potential energy. This behaviour is reminiscent of the findings of Aussillous \& Qu\'{e}r\'{e} \cite{aussillous2006properties} for non-wetting drops on rigid substrates: the upper portion of the drop is essentially flat. An important distinction from the rigid substrate case, however, is that a bulge is formed below the TPCL. This reduces the radial length scale of the drop (for a given volume) by acting as a reservoir, in which the drop volume can be stored. We shall analyse this shape quantitatively later in this section, seeking to determine the radial position of the TPCL, $r_c$.

\subsubsection{Rigid substrate approximation}\label{rigid}

To obtain a first approximation for $r_c$ we imagine that the substrate is rigid. Intuitively, we expect this approximation to be valid provided that $\Gdv\ll 1$, since then the drop will prefer  to minimize its gravitational potential energy by spreading horizontally instead of deforming the substrate vertically (see e.g.~fig.~\ref{fgr:transition}(\emph{b})). 

We begin by following the scaling argument by Aussillous \& Qu\'{e}r\'{e} \cite{aussillous2006properties} for deformable drops on rigid substrates. Such drops adopt a puddle shape, with a constant thickness $h$ over most of their length. To determine $h$, we balance the hydrostatic pressure along the central plane of the droplet, $\rhod gh/2$ with the curvature induced pressure from the rounded edges of the droplet, $\gdv/(h/2)$.
 In this approximation, the height of the puddle satisfies
\begin{equation}
h\approx 2\sqrt{\frac{\Gdv}{D}}\lc=2\lc^D
\label{puddle_h}
\end{equation}
where $\lc^D=\sqrt{\Gdv/D}\lc$ is the capillary length of the drop. The position of the contact line, $r_c$, may then be estimated from the volume constraint, $V \approx \pi r_c^2 h$, to give
\begin{equation}
R_c=\frac{r_c}{\lc}\approx\sqrt{\frac{2}{3}} \left(\frac{D}{\Gdv}\right)^{1/4}\Bo^{3/4}.
\label{puddle_rc}
\end{equation} 
We expect that this scaling may hold for $\lc^D\ll r_c \ll \lc$: the drop should be large enough to be a puddle, but small enough to remain afloat. We also note that the droplet radius given in \eqref{puddle_rc} scales as $\Bo^{3/4}$, which is significantly different from the $r_c\propto\Bo^{1/2}$ and $r_c\propto\Bo$ scalings that we saw for relatively undeformable droplets in \S\ref{sec:undeform}.

Numerical results, presented in fig.~\ref{fgr:deform_theory} show that the rigid substrate approximation, \eqref{puddle_rc}, provides an upper bound on $r_c$ for a given value of $\Bo$. However, we also note that there is a significant discrepancy, particularly as $\Bo$ increases: \eqref{puddle_rc} over-estimates the lateral size of the droplet since a significant amount of volume is stored in the bulged, central part of the droplet (thereby decreasing $r_c$).  To understand and account for this discrepancy, we construct a refined analytical model of the drop that takes this bulging into consideration. 

\subsubsection{Refined analytical model}\label{refine}

\begin{figure}[t!]
\centering
\includegraphics[width=0.8\columnwidth]{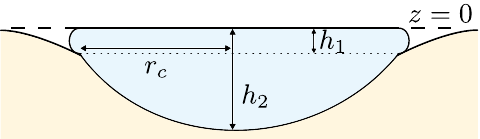}
\caption{Schematic diagram of the refined analytical model.}
\label{fgr:deform_diagram}
\end{figure}

We begin by modelling the bulge as a spherical cap with radius of curvature $\mathcal{R}$ (since the drop is small compared to the capillary length of the substrate, this seems reasonable). We also assume that the drop-vapour interface has negligible curvature and is level with the height of an undeformed substrate, so that $\zmax\approx0$. As in fig.~\ref{fgr:deform_diagram} we specify two heights for the droplet:  $h_1$ is  the height between the top apex and the TPCL (so that the vertical position of the TPCL is $z\approx-h_1$) and $h_2$ is the maximal thickness of the drop.

At the base of the drop, the Laplace--Young equation for the drop--liquid interface gives 
\begin{equation}
(\rhod-\rhol)gh_2= \frac{2\gdl}{\mathcal{R}}
\end{equation}
and hence
\begin{equation}
h_2= \frac{2\Gdl}{D-1}\frac{\lc^2}{\mathcal{R}}\text{.}
\label{h2}
\end{equation}
If we then assume that the bulge is relatively small, and may be approximated by a parabola, we may write
\begin{equation}
h_2-h_1= \frac{r_c^2}{2\mathcal{R}}\text{.}
\label{refine_parabola}
\end{equation} 

The balance of pressures that led to  \eqref{puddle_h} applies equally to $h_1$, while \eqref{h2} determined $h_2$. We may therefore substitute these results into  \eqref{refine_parabola} to write
\begin{eqnarray}
\mathcal{R}=\frac{\Gdl}{D-1}\frac{\lc^2}{\lc^D}\left[1-\frac{D-1}{4\Gdl}\left(\frac{r_c}{\lc}\right)^2\right]\text{.}
\label{R_expression1}
\end{eqnarray}
We rescale $\mathcal{R}$ by $\lc$ and rewrite the right-hand side of \eqref{R_expression1} in terms of $\lc$
	\begin{eqnarray}
	\frac{\mathcal{R}}{\lc}=\frac{\Gdl}{D-1}\sqrt{\frac{D}{\Gdv}}\left[1-\frac{D-1}{4\Gdl}\left(\frac{r_c}{\lc}\right)^2\right]\text{.}
	\label{R_expression2}
	\end{eqnarray}
Since we are assuming that $r/\lc\ll1$ and $\Gdl=1+\Gdv$ (for non-wetting drops), this may be simplified to give
	\begin{equation}
	\frac{\mathcal{R}}{\lc}\approx \frac{1}{D-1}\sqrt{\frac{D}{\Gdv}}{.}
	\label{R_expansion}
	\end{equation}
Now, in order to find a relationship between $r_c$ and $V$, we must consider the conservation of volume. Combining the cylindrical approximation for the volume of the flat cap with the volume of the paraboloid, we have 
\begin{eqnarray}
\frac{4\pi}{3}\Bo^{3/2}\lc^3&\approx &\pi h_1r_c^2+\pi r_c^2 (h_2-h_1)/2\label{refine_volume_1}\\
&=&2\pi \lc^D r_c^2 +\frac{\pi}{4\mathcal{R}}r_c^4,
\label{refine_volume}
\end{eqnarray}
using \eqref{refine_parabola}. Taking the leading order approximation of $\mathcal{R}$ from \eqref{R_expansion},  independent of $r_c$, we find a quadratic equation for $r_c^2$:
\begin{equation}
\frac{1}{8}(D-1)\left(\frac{r_c}{\lc}\right)^4+\left(\frac{r_c}{\lc}\right)^2-\frac{2}{3}\frac{\Bo^{3/2}\lc}{\lcd}=0\text{,}
\end{equation}
and hence
\begin{equation}
\frac{r_c}{\lc}=R_c=\frac{2}{\sqrt{D-1}}\left\{\left[1+\frac{D-1}{3}\frac{\Bo^{3/2}\lc}{\lc^D}\right]^{1/2}-1\right\}^{1/2}\text{.}
\label{refine_final}
\end{equation}
At first sight, $r_c$ is now expressed in terms both $\lc$ and $\lc^D$, which is different from results we have had thus far. However, before we discuss the physical significance of this expression, we first establish its connection with the rigid substrate model presented in \S\ref{rigid}. This rigid substrate limit should be recovered in the limit $\Bo\lll1$; a Taylor expansion of \eqref{refine_final} clearly shows that this expectation is correct.

Continuing this Taylor expansion, we find that the next order correction is negative, corresponding to a decrease in the contact radius $r_c$ compared to the rigid substrate model. This therefore captures something of the `bulging' of the droplet that is neglected by the rigid substrate model. Furthermore, the size of the correction also increases with $\Bo$,  consistent with the preliminary numerical results presented in fig.~\ref{fgr:transition}(\emph{a}).

\begin{figure}[t!]
\centering
\includegraphics[width=0.7\columnwidth]{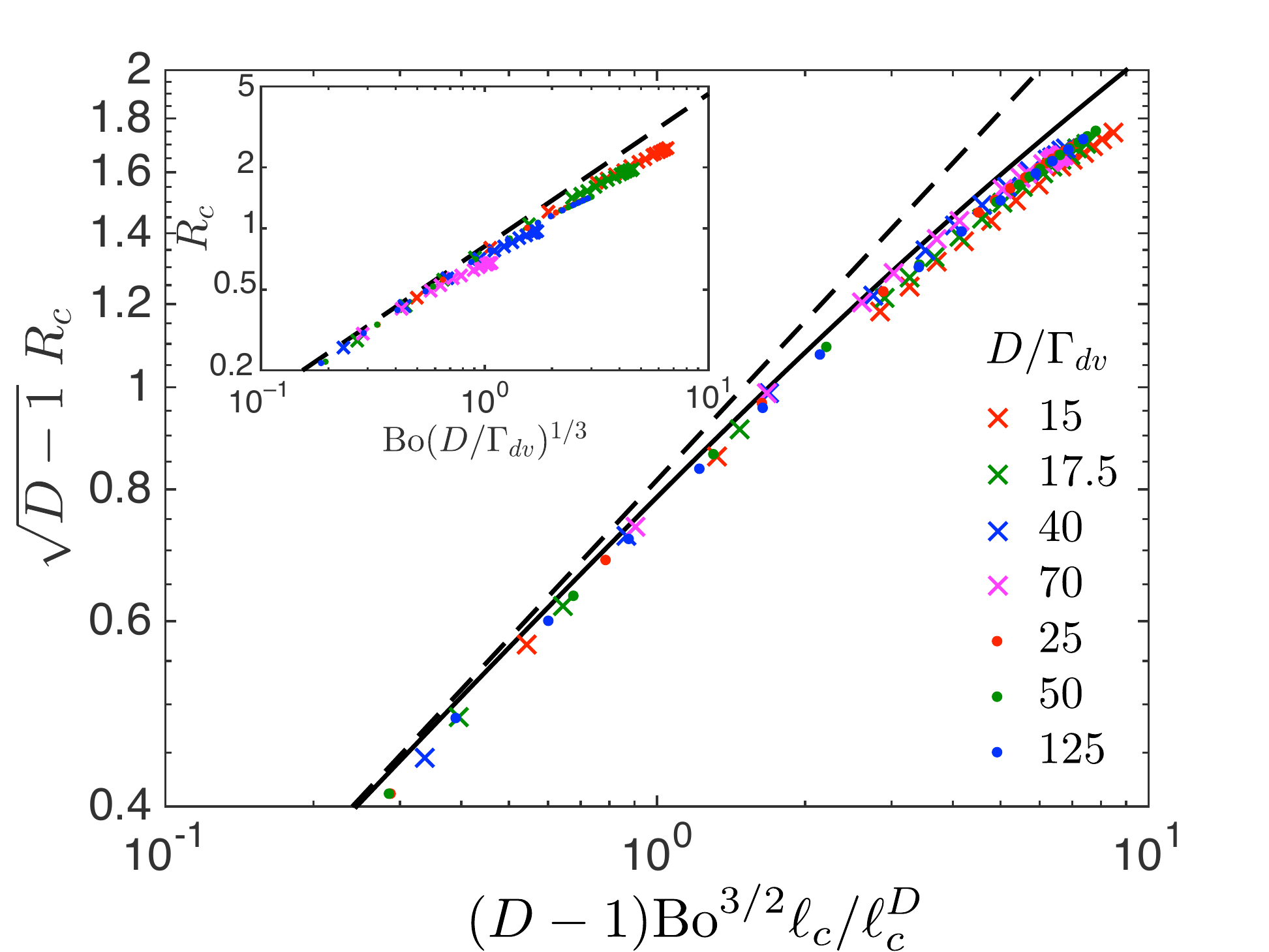}
\caption{Scaling behaviour of the contact radius as a function of drop volume in the highly deformable limit. Crosses ($\times$) denote numerical results with $\Gdv=0.1$ and $D$ varying while bullets ($\bullet$) denote results with $D=2.5$ and $\Gdv$ varying. The black solid curve indicates the theoretical prediction of the refined model, \eqref{refine_final}, while the dashed line indicates the scaling prediction for rigid substrates, \eqref{puddle_rc}, which is the generalization of a previous result \cite{aussillous2006properties}. The inset gives the same data with the nondimensionalization suggested by \eqref{puddle_rc},  highlighting that the collapse in the main figure is a considerable improvement.}
\label{fgr:deform_theory}
\end{figure}

We compare the predictions of the two models (the rigid substrate and the refined models) with numerical simulations in fig.~\ref{fgr:deform_theory}. We see that when plotting the data in the way suggested by the rigid substrate model (see inset of fig.~\ref{fgr:deform_theory}) the agreement is qualitatively good, but that there is considerable scatter in the data. However, the rescaling suggested by the refined model \eqref{refine_final}, shown as the main figure in fig.~\ref{fgr:deform_theory}, show first that the data collapse significantly better when plotted in this way, and, second, that the trend of \eqref{refine_final} gives a much better quantitative account of the numerical results. The discrepancy that remains is presumably a result of the approximation that $\zmax\approx0$ for drops with $\Bo\lesssim1$. Finally, we note that in the limit of very small Bond numbers, even drops with $\Gdv\ll1$ eventually become relatively undeformable and return to the limit considered in \S\ref{sec:undeform}.

\section{Conclusions}

We have considered the flotation of liquid drops on the surface of another liquid, focusing in particular on the case where the droplet is close to being perfectly non-wetting. This is a scenario that is common in a number of applications, including liquid marbles, Leidenfrost droplets and droplets on a vibrating bath of the same liquid. 

We presented numerical and analytical results for a range of different parameters but focussed in particular on two sets of behaviour corresponding to relatively undeformable droplets, as well as very deformable droplets. In the first case, we showed that the floating state can be understood using simple ideas of force balance. In the limit of small droplets, this force balance can be simplified to the extent that we are able to present analytical results for the state in which the droplet floats. In the second case (very deformable droplets), we found that the classic picture of a flat pancake or puddle is significantly modified because of the deformability of the liquid interface on which the droplet floats. We provided a refined analytical model that accounts for this deformability and provides improved agreement with numerical results. This model also emphasizes that the problem involves multiple capillary lengths, $\lc$ and $\lcd$, and that it is a combination of the two that determines the behaviour of the system. We also demonstrated that some features of the quasi-static levitation of  Leidenfrost drops are quantitatively captured by modelling such drops as static, perfectly non-wetting drops. This provides new insight into their floating state, without necessarily developing detailed hydrodynamic models of levitation.

A key finding of our analysis is that the properties of the floating state for relatively undeformable drops are surprisingly sensitive to the relative tensions of the drop--vapour and drop--liquid interfaces. In particular,  the meniscus inclination, $\theta_3$, and contact line radius exhibit different size-dependent scalings depending on these differences. We suggest that the measurement of the floating state of such droplets may be a sensitive, non-invasive assay for the measurement of interfacial tensions close to perfect non-wetting. This possibility seems particularly relevant for the particle--coated interfaces that occur in liquid marbles, but are also used to stabilize emulsions \cite{Binks2006book}. 


\section*{Acknowledgements}

The research leading to these results has received funding from the European Research Council under the European Union's Horizon 2020 Programme / ERC Grant Agreement no.~637334 (DV).

\section*{Appendix A: Dimensionless problem and numerical solution}

\subsection*{Dimensionless problem}

The non-dimensionalization reported in \S\ref{sec:nondim} transforms the problem \eqref{rz1}--\eqref{twodim} to:
\begin{eqnarray}
\frac{\upd R_i}{\upd S}&=&\cos\phi_i, \label{one}\\
\frac{\upd Z_i}{\upd S}&=&\sin\phi_i, \label{two}\\
\frac{\upd\phi_i}{\upd S}&=&\begin{cases}\frac{D}{\Gdv}Z_i-\frac{\sin{\phi_i}}{R_i}-\frac{D}{\Gdv}Z_{0} & \mbox{for}\ i=1\\
\frac{1-D}{\Gdl}Z_i-\frac{\sin{\phi_i}}{R_i}+\frac{D}{\Gdl}Z_{0}& \mbox{for}\ i=2\\
Z_i-\frac{\sin{\phi_i}}{R_i}& \mbox{for}\ i=3\text{.}\end{cases} \label{three}
\end{eqnarray}

Using $S=0$ to denote the centre of the droplets, we have the symmetry boundary conditions
\begin{equation}
R_{1,2}(0)=0,\quad \phi_{1,2}(0)=0.
\end{equation}
The position of the  TPCL is denoted by $S_{1,2}^{(c)}$ in arc length coordinates. At this point the appropriate boundary conditions are continuity with the expressions for the outer meniscus. Denoting the coordinates of the TPCL by $(R_c,Z_c)$ we find that
\begin{center}
\begin{tabular}{l l l} 
$R_{1,2}(S_{1,2}^{(c)})=R_c$, &$Z_{1,2}(S_{1,2}^{(c)})=Z_c$, &$\phi_i(S_i^{(c)})=(-1)^i\theta_i$  \\ 
 $R_3(0)=R_c$, &$Z_3(0)=Z_c$, & $\phi_3(0)=\theta_3$\text{.} \\
\end{tabular}
\end{center}
Finally, we apply the decay condition for the outer meniscus
\begin{equation}
Z_3(\infty)=0.
\end{equation}
We therefore have a total of 14 boundary conditions for the 9 differential equations \eqref{one}-\eqref{three}. This apparent discrepancy is explained by the system having five additional quantities $Z_{0}$, $R_c$, $\theta_3$, $S_1^{(c)}$ and $S_2^{(c)}$ that are not known \emph{a priori}. (The contact angles $\theta_1$ and $\theta_2$ are determined in terms of $\theta_3$ and the interfacial energies $\Gdv$ and $\Gdl$ by the Neumann relations \eqref{neumann1}-\eqref{neumann2}; similarly, $Z_c$ is determined by the outer meniscus, once $\theta_3$ and $R_c$ are given.)

\subsection*{Numerical scheme} \label{numerics}

Following ref.~\cite{phan2012can}, and for convenience, we impose $D$ and $\theta_3$ and compute the drop volume $V$ that would give this angle.  However, the height and radial position of the TPCL are unknown. We therefore solve for the upper and lower drop surfaces using a shooting method that terminates when both interfaces meet at the TPCL (this determines the values of $S_{1,2}^{(c)}$ since $\phi_1(S_1^{(c)})=-\theta_1$ and $\phi_2(S_2^{(c)})=\theta_2$, are given). This stage of the numerics is completed  via the EventLocator function in \textit{Mathematica}. In practice, a change of variables in \eqref{one}-\eqref{three} facilitates the solution, by removing the coordinate singularity near the drop apices; we let $\phi_{1,2}(S)=S\psi_{1,2}(S)$ and $r=S\eta_{1,2}(S)$. 

To solve for the outer meniscus, we  use the MATLAB boundary value problem solver \texttt{bvp4c}. This outer meniscus problem is decoupled from the full problem once $R_c$ is determined from the first step. Finally, all three interfaces are matched. This hybrid method (using both Mathematica and MATLAB) is easily achieved using MATLink.

\subsection*{Verification of code}
In the absence of previous analytical results with which to compare our numerical calculations, we verify numerically that the drop satisfies a global force balance condition: the generalized Archimedes' Principle \cite{keller1998,vella2015floating} shows that the restoring force on the drop equals the total weight of liquid displaced (including that displaced in the menisci). For practical purposes, it is often simpler to think of the weight of liquid that is displaced by the wetted portion of the drop, i.e.~$\rhol g\tilde{V}$, and the weight displaced in the menisci separately since the latter can be equated to the vertical force from surface tension acting along the contact line \cite{vella2015floating}. We therefore have that the global force balance reads
\begin{equation}
\rho_d gV=\rho_lg\tilde{V}+2\pi\gamma_{lv} r_c \sin\theta_3.
\label{fb}
\end{equation} We check that this is satisfied to within $0.01\%$  in all of the numerical results presented here. The agreement between numerical and asymptotic results appropriate to small droplets gives a final check on the validity of the numerical results.


\end{document}